# Estimation of Poisson Autoregressive Model for Multiple Time Series

*Paolo Victor T. Redondo*
*Graduate Student, School of Statistics, University of the Philippines Diliman*
ptredondo@up.edu.ph
*Joseph Ryan G. Lansangan*
*Professor, School of Statistics, University of the Philippines Diliman*
jglansangan@up.edu.ph
*Erniel B. Barrios*
*Professor, School of Statistics, University of the Philippines Diliman*
ebbarrios@up.edu.ph
ORCID:0000-0003-4209-2858

**ABSTRACT**

A Poisson autoregressive (PAR) model accounting for discreteness and autocorrelation of count time series data is typically estimated in the state-space modelling framework through extended Kalman filter. However, because of the complex dependencies in count time series, estimation becomes more challenging. PAR is viewed as an additive model and estimated using a hybrid of cubic smoothing splines and maximum likelihood estimation (MLE) in the backfitting framework. Simulation studies show that this estimation method is comparable or better than PAR estimated in the state-space context, especially with larger count values. However, as [2] formulated PAR for stationary counts, both estimation procedures underestimate parameters in nearly nonstationary models. The flexibility of the additive model has two benefits though: robust estimation in the presence of temporary structural change, and; viability to integrate PAR model into a more complex model structure. We further generalized the PAR(p) model into multiple time series of counts and illustrated with indicators in the financial markets.

1.  **INTRODUCTION**

Importance of analyzing count data time series is well distinguished in many areas. In political communication studies, [8] analyzed count data on themes about church, energy policy and homosexuality. In financial markets, Liesenfeld et al. (2006) developed a dynamic model for integer counts capturing fundamental properties of discrete price movements. Incidence of poliomyelitis [6], daily admissions for asthma [6], [10], dependence of the occurrence of infectious disease on weather[9] are some examples of count data that exhibits time dependencies in medicine and epidemiology.

Time series of count of events are often characterized by stylized facts like discreteness, non-normality, possible overdispersion, serial correlation, non-stationarity and mean-reversion that have repercussions on modelling techniques. These characteristics often violates typical assumptions in standard analyses resulting to invalid inferences [2], [8]. To account for these behaviors of the count series, [2] developed the Poisson Autoregressive PAR(p) model; a dynamic State-space model that assumes a Poisson data generating process as a measurement equation in the state-space methodology.

PAR(p) model is estimated in the state-space framework, specifically with the Kalman Filter which extends the system dynamics to nonlinear functions. Kalman filter updates knowledge of the state variable recursively when a new data point becomes available, i.e., knowing the conditional distribution of the state given all available observations in its history. Time series data in longer horizon usually causes overparametrization in estimation leading to inadequate inferences. Moreover, since the state/transition equation of a PAR(p) model concerns the dynamic mean, the mean and variance level of the count time series greatly affects the MLE-based estimates in the state-space formulation often resulting to inadmissibility of the estimates produced.

Furthermore, PAR(p) model may generate count outcomes that may deviate far away from large dynamic means, resulting to proliferation of measurement error. Whenever these measurement errors are much larger than model stochasticity, [1] demonstrated, through simulation studies, that State-space models (even simple linear Gaussian state-space models), estimated through MLE, suffers from overparametrization of the observation and state models.

To avoid issues associated with Kalman Filtering, instead of viewing the PAR(p) model in a state-space perspective, we interpret the PAR(p) model as an additive model and propose a hybrid estimation procedure imbedded into the backfitting algorithm. Methods of analysis of count time series data and the advantages of the PAR(p) model is discussed in Section 2. We discuss the alternative framework of analysis for count time series along with the hybrid estimation procedure in Section 3. A simulation study to compare the performance of the hybrid estimation and extended Kalman-filter is presented in Section 4. We also compare the performance of the hybrid estimation and extended Kalman-filter in the presence of temporary structural change in Section 5. The PAR(p) model is generalized into a multiple count time series in Section 6 and is used with indicators in the financial market in Section 7. Some important results and implications are presented in Section 8.

## 2. ANALYSIS OF COUNT TIME SERIES

Temporal dependence that causes overdispersion of counts is often disregarded in modeling count data when Negative Binomial model is used since it also account for overdispersion. Count data may also be transformed (e.g., logarithm) to satisfy certain assumptions of a Gaussian ARIMA model, after the event count properties of the data is ignored. Both techniques however, usually fail to adequately capture the dynamics of the data and/or the data generating process that gives rise to count outcomes [2], [8].

In capturing salient features of the count time series data, two classes of modeling approaches were proposed, the parameter-driven and observation-driven methods. In the former, a latent dynamic process governs the conditional mean function while the latter specifies its dependence structure by incorporating lagged values of the observed counts directly into the mean function of the model [10]. Parameter-driven specification allows for easy derivation of model properties and interpretability of parameter estimates in exchange of having difficulties in estimation and prediction due to complicated likelihood functions. In contrast, easy to calculate likelihood functions and straightforward predictions are generally evident in observation-driven methods at the expense of difficulties in characterizing model properties as well as the non-interpretable parameter estimates.

To address issues for both approaches, i.e., having well-defined likelihood functions, established model properties and interpretable parameter estimates, [3] and [2] postulated two classes of models for dynamic count event time series processes; Poisson exponentially weighted moving average (PEWMA) and Poisson Autoregressive (PAR) models, both of which belong to the state-space modeling framework.

The PEWMA model can be summarized by the following equations.

$$Y_t \sim \text{Poisson}(m_t) \text{ i.e.}, P(Y_t = y|m_t) = \frac{m_t^y e^{-m_t}}{y!} \text{ (measurement)} \quad (1)$$

$$m_t = \exp(r_t)\mu_{t-1}\eta_t \text{ where } \eta_t \sim \beta(\omega a_{t-1}, (1-\omega)a_{t-1}) \text{ (transition)} \quad (2)$$

$$m_t|Y_{t-1} \sim Gamma(a_{t-1}, b_{t-1}), m_{t-1} > 0 \text{ (conjugate prior)} \quad (3)$$

Simulation studies provided evidence on the superiority of PEWMA in explaining a dynamic non-stationary count processes, but, fails to capture the time dependent structure whenever the series is stationary and mean-reverting [3].

The Poisson Autoregressive PAR(p) model is specified by the following system of equations:

$$Y_t \sim \text{Poisson}(m_t) \text{ i.e.}, P(Y_t = y|m_t) = \frac{m_t^y e^{-m_t}}{y!} \text{ (measurement)} \quad (4)$$

$$m_t = \sum_{i=1}^p \rho_i Y_{t-i} + \left(1 - \sum_{i=1}^p \rho_i\right)\exp(\delta_0 + X_t`\boldsymbol{\delta}) \text{ (transition)} \quad (5)$$

$$m_t|Y_{t-1} \sim Gamma(\sigma_t m_{t-1}, \sigma_{t-1}), m_{t-1} > 0, \sigma_{t-1} > 0 \text{ (conjugate prior)} \quad (6)$$

where $Y_t$ – count response at time t

$m_t$ – dynamic mean of the Poisson data generating process at time t

$\rho_i$ – autoregressive parameters $i = 1,2,...,p$

$X_t$ – vector of covariates at time t

$\boldsymbol{\delta}$ – vector of coefficients of each covariate

$\delta_0$ – mean-reversion point of the count series

$m_{t-1} = E(y_t|Y_{t-1})$

$\sigma_{t-1} = Var(y_t|Y_{t-1})$

For both PEWMA and PAR, the use of a Poisson data generating process as the measurement equation provides a natural accounting of the discreteness of count data. For PAR, the dynamic mean specification imposes stationarity and mean-reversion while it also account for the contribution of the exogenous covariates and temporal dependencies. Also, the conjugate gamma prior makes the one-step ahead forecast distribution to be negative binomial and allows for overdispersion. [8] noted that the specifications of the model, allowing for a dynamic causal effect (while imposing condition on stationarity) and overdispersion, addresses all the attributes of time-dependent media count data. Further characterization and estimation of PEWMA and PAR, along with interpretation of parameters, are further discussed in [2].

Admissibility of resulting estimates depends on the mean level and variance of the series [2]. Since large dynamic mean translates to large measurement errors as demonstrated by [1], estimation of the PAR(p) model in the state-space context suffers from severe parameter- and state-estimation problems when dealing with large count time series data. Thus, we view the PAR(p) model as additive model and propose an alternative estimation procedure from the hybrid of cubic smoothing splines and maximum likelihood estimation (MLE) in the backfitting framework.

## 3. ESTIMATION OF ADDITIVE POISSON AUTOREGRESSION

To exhibit additivity of $PAR(p)$ model, consider the transition equation whose mean or state variable evolves through a stationary $AR(p)$ process with exogenous regressors $X_t$ given in Equation (7).

$$m_t = \sum_{i=1}^{p} \rho_i y_{t-i} + (1 - \sum_{i=1}^{p} \rho_i) exp(\delta_0 + X_t`\delta) \tag{7}$$

Rewrite the term $(1 - \sum_{i=1}^{p} \rho_i)$ as $exp\left(ln(1 - \sum_{i=1}^{p} \rho_i)\right)$ in Equation (8) and note that this term is independent of the covariate effect (Equation 9) and can be included in the mean reverting point of the count series (Equation 10).

$$m_t = \sum_{i=1}^{p} \rho_i y_{t-i} + exp\left(ln(1 - \sum_{i=1}^{p} \rho_i)\right) exp(\delta_0 + X_t`\delta) \tag{8}$$

$$m_t = \sum_{i=1}^{p} \rho_i y_{t-i} + exp(ln(1 - \sum_{i=1}^{p} \rho_i) + \delta_0 + X_t`\delta) \tag{9}$$

$$m_t = \sum_{i=1}^{p} \rho_i y_{t-i} + exp(\delta_0^* + X_t`\delta) \tag{10}$$

where $\delta_0^* = ln(1 - \sum_{i=1}^{p} \rho_i) + \delta_0$

$$m_t = f_1(Y_{t-1}, \dots, Y_{t-p}) + f_2(X_t) \tag{11}$$

where $f_1(Y_{t-1}, \dots, Y_{t-p}) = \sum_{i=1}^{p} \rho_i y_{t-i}$ and $f_2(X_t) = exp(\delta_0^* + X_t`\delta)$

Clearly, the transition equation exhibits additivity in Equation (11) which is crucial for the hybrid estimation in the backfitting framework. In addition, the first component $f_1(Y_{t-1}, \dots, Y_{t-p})$ is just a function of lagged values of the count series, hence, a natural estimator for $\rho_i$ is provided by the first derivative of the first component with respect to $Y_{t-i}$, i.e., $\hat{\rho}_i = \frac{\partial f_1(Y_{t-1}, \dots, Y_{t-p})}{\partial Y_{t-i}}$. The second component, $f_2(X_t)$ is similar with the mean function of the usual

Poisson regression. Thus, a maximum likelihood estimator (MLE) may be used to estimate the covariate effects $\boldsymbol{\delta}$. We incorporate these two estimators then in the backfitting algorithm.

*Hybrid Estimation of PAR(p) Model*

For any random variable following a certain distribution ($X \sim F_X$), a single observation $X_t$ is an unbiased estimator for its mean [$E(X)$]. At time $t$, suppose the observed outcome of the Poisson data generating process is $Y_t$. Since $Y_t$ comes from a Poisson process with mean $m_t$, $Y_t$ is an unbiased estimator of $m_t$. Thus, the transition equation can be represented as

$$\widehat{m}_t = Y_t = \sum_{i=1}^{p} \rho_i Y_{t-i} + \left(1 - \sum_{i=1}^{p} \rho_i\right) \exp(\delta_0 + \boldsymbol{X}_t`\boldsymbol{\delta}) \qquad (12)$$

We estimate Equation (12) as follows:

*Step1*: Ignore the autoregressive term and fit $Y_t$ with covariates using Poisson Regression to obtain $\hat{\delta}_o^{(0)}$ and $\hat{\boldsymbol{\delta}}^{(0)}$.

*Step2*: Compute residuals $R_t^{(1)} = Y_t - \exp(\hat{\delta}_o^{(0)} + \boldsymbol{X}_t`\hat{\boldsymbol{\delta}}^{(0)})$.

Note that the estimates from Step1 are biased because the misspecification error when the first term in Equation (12) is ignored. Also, $\hat{\delta}_o^{(0)}$ is the estimate for the $\delta_0^*$ in Equation (10) and not the mean reverting point $\delta_0$. This will properly scaled upon convergence of this estimation algorithm.

*Step3*: Fit $R_t^{(1)}$ as a cubic smoothing spline function of $R_{t-i}^{(1)}, i = 1,2, \dots, p$ to generate an estimator of the first component denoted by $\hat{f}_1(Y_{t-1}, \dots, Y_{t-p})$. The estimator for the autoregressive parameter is computed from the slope $\hat{\rho}_i^{(0)} =$

$\left(\sum_{t=2}^{T} \frac{\partial \widehat{f_1}(Y_{t-1},\ldots,Y_{t-p})}{\partial R_{t-i}^{(1)}}\right) / (T-1)$, i.e., the mean of the analytical first derivative of $\widehat{f_1}(Y_{t-1},\ldots,Y_{t-p})$ with respect to $R_{t-i}^{(1)}$ evaluated at every time point $t$.

*Step4*: Define new residuals as $R_t^{(2)} = Y_t - \sum_{i=1}^{p} \hat{\rho}_i^{(0)} Y_{t-1}$.

For $j = 1, 2, 3, \ldots$ where $j$ is the index of the iteration and initiate the residuals $R_t^{(2j)} = R_t^{(2)}$.

*Step5*: Fit $R_t^{(2j)}$ with covariates using Poisson regression to obtain $\hat{\delta}_o^{(j)}$ and $\widehat{\boldsymbol{\delta}}^{(j)}$.

Note that the misspecification error accrued in *Step1* is now minimized after the adjustment with the estimated autoregressive component.

*Step6*: Define new residual as $R_t^{(3j)} = Y_t - \exp(\hat{\delta}_o^{(j)} + \boldsymbol{X}_t`\widehat{\boldsymbol{\delta}}^{(j)})$.

*Step7*: Fit $R_t^{(3j)}$ as a cubic smoothing spline function of $R_{t-i}^{(3j)}, i = 1,2,\ldots,p$ to generate an estimator of the autoregressive component denoted by $\widehat{f_1}(Y_{t-1},\ldots,Y_{t-p})$. The estimator for the autoregressive parameter is computed from $\hat{\rho}_i^{(j)} = \left(\sum_{t=2}^{T} \frac{\partial \widehat{f_1}(Y_{t-1},\ldots,Y_{t-p})}{\partial R_{t-i}^{(3j)}}\right) / (T-1)$.

*Step8*: Define new residual as $R_t^{(2j+1)} = Y_t - \sum_{i=1}^{p} \hat{\rho}_i^{(j)} Y_{t-1}$.

Iterate from *Step5*, defining residuals using the updated estimates for $\rho_i$ and $\boldsymbol{\delta}$ in every iteration until the convergence criterion is achieved, e.g., changes in all parameter estimates are less than 0.0001%.

## 4. SIMULATION STUDIES

To assess the performance of the estimation procedure over the extended Kalman-filter, we conducted simulation studies with settings summarized in Table 1.

**Table 1. Summary of the Simulation Settings**

| Autoregressive Parameter ($\rho$) | Covariate Coefficient ($\delta$) | Covariate Structure | Series Length |
|---|---|---|---|
| 0.2 | 0.25 | Uniform(0,1) | 100 |
| 0.6 | 0.5 | Normal(0,1) | 200 |
| 0.95 | | Poisson(5) | 500 |

We simulated count time series from a $PAR(1)$ data generating process with a single covariate and a fixed mean reverting value equal to 100. This will ensure that the data is not zero-inflated since this will cause another issue in estimating the model. We then compare results of the hybrid estimation with the extended Kalman-filter. For each of the scenarios (54), we simulated 200 replicates, the model is estimated in each replicate using both procedures, average relative bias and predictive ability (measured through the mean absolute percentage error (MAPE) and root mean square error (rMSE) are used as basis for the comparison.

*Covariate: Normal(0,1)*

With the covariate generated from $N(0,1)$, the estimates (averaged from 200 replicates) and standard errors of the autoregressive parameter and covariate effect are given in Tables 2 and 3, respectively. The indicators of predictive ability and convergence rates are presented in Table 4.

Given a normally distributed covariate, both estimation methods have almost certain convergence except for cases of near non-stationarity (see Table 4). For fairly stationary count time series ($\rho = 0.2, 0.6$), the proposed estimation procedure works comparably with the extended Kalman-filter in terms of standard error and relative bias of the autoregressive and covariate effect (see Tables 2 and 3). Even the predictive ability represented by MAPE and rMSE (see Table 4) are comparable for the hybrid estimation and the extended Kalman-filter. Increasing length of the time series leads to minimal decline in standard errors and bias of both the autoregressive parameters and the covariate effects. Predictive abilities and convergence rates of the two methods are fairly robust to length of the time series, autoregressive parameter, and the covariate effects.

Stationarity implies that the series does not have any persistent effect on its own. This translates to having minimal dependencies which makes the backfitting algorithm works in the time series context [5]. However, as the time series approaches nonstationarity ($\rho = 0.95$), both estimation procedures fail to estimate the parameters of the time series well.

From the transition equation of the PAR(p) model in Equation (5), whenever $\rho \to 1$, the contribution of the covariates to the dynamic mean diminishes because of the coefficient $(1 - \rho)$ which imposes the challenge for both estimation procedures. While both methods estimated the autoregressive parameter $\rho$ (see Table 2) fairly well, their respective estimates for the covariate effect $\delta$ suffers. Despite having relatively low standard errors, estimates for the covariate effect of the hybrid procedure are biased. On the other hand, estimates from the extended Kalman-filter may exhibit lower bias but has large standard errors (see Table 2). Results for near-nonstationary cases are expected since PAR(p) model was formulated with the assumption that the time series is stationary.

**Table 2. Estimates for $\rho$, Standard Error and Relative Bias with Normal Covariate**

| $\rho$ | $\delta$ | Series Length | Hybrid Estimation | | | Extended Kalman-Filter | | |
|---|---|---|---|---|---|---|---|---|
| | | | Estimate | St. Err. | Relative Bias | Estimate | St. Err. | Relative Bias |
| 0.2 | 0.25 | 100 | 0.1968 | 0.0444 | 17.57 | 0.1980 | 0.0416 | 16.44 |
| 0.2 | 0.25 | 200 | 0.1983 | 0.0324 | 12.33 | 0.1995 | 0.0299 | 11.50 |
| 0.2 | 0.25 | 500 | 0.1984 | 0.0200 | 8.07 | 0.1992 | 0.0188 | 7.49 |
| 0.2 | 0.5 | 100 | 0.1998 | 0.0275 | 11.00 | 0.1998 | 0.0201 | 8.07 |
| 0.2 | 0.5 | 200 | 0.2005 | 0.0199 | 7.92 | 0.2011 | 0.0144 | 5.64 |
| 0.2 | 0.5 | 500 | 0.2000 | 0.0123 | 4.87 | 0.2004 | 0.0095 | 3.79 |
| 0.6 | 0.25 | 100 | 0.5788 | 0.0539 | 7.50 | 0.5855 | 0.0522 | 7.03 |
| 0.6 | 0.25 | 200 | 0.5911 | 0.0401 | 5.37 | 0.5942 | 0.0388 | 5.18 |
| 0.6 | 0.25 | 500 | 0.5956 | 0.0245 | 3.08 | 0.5973 | 0.0247 | 3.21 |
| 0.6 | 0.5 | 100 | 0.5932 | 0.0364 | 4.85 | 0.5979 | 0.0282 | 3.85 |
| 0.6 | 0.5 | 200 | 0.5972 | 0.0283 | 3.51 | 0.6008 | 0.0220 | 2.81 |
| 0.6 | 0.5 | 500 | 0.5992 | 0.0154 | 2.05 | 0.6005 | 0.0133 | 1.80 |
| 0.95 | 0.25 | 100 | 0.8739 | 0.0543 | 8.10 | 0.9025 | 0.0547 | 5.56 |
| 0.95 | 0.25 | 200 | 0.9147 | 0.0284 | 3.89 | 0.9301 | 0.0268 | 2.71 |
| 0.95 | 0.25 | 500 | 0.9338 | 0.0171 | 1.90 | 0.9405 | 0.0162 | 1.47 |
| 0.95 | 0.5 | 100 | 0.8806 | 0.0495 | 7.41 | 0.9118 | 0.0503 | 4.72 |
| 0.95 | 0.5 | 200 | 0.9160 | 0.0282 | 3.70 | 0.9361 | 0.0259 | 2.33 |
| 0.95 | 0.5 | 500 | 0.9351 | 0.0171 | 1.76 | 0.9434 | 0.0163 | 1.28 |

**Table 3. Estimates for Covariate Effect $\delta$, Standard Error, and Relative Bias with Normal Covariate**

| $\rho$ | $\delta$ | Series Length | Hybrid Estimation | | | Extended Kalman-Filter | | |
|---|---|---|---|---|---|---|---|---|
| | | | Estimate | St. Err. | Relative Bias | Estimate | St. Err. | Relative Bias |
| 0.2 | 0.25 | 100 | 0.2481 | 0.0194 | 6.20 | 0.2486 | 0.0185 | 5.79 |
| 0.2 | 0.25 | 200 | 0.2485 | 0.0131 | 4.05 | 0.2489 | 0.0121 | 3.87 |
| 0.2 | 0.25 | 500 | 0.2488 | 0.0080 | 2.60 | 0.2491 | 0.0076 | 2.49 |
| 0.2 | 0.5 | 100 | 0.4995 | 0.0210 | 3.36 | 0.4993 | 0.0168 | 2.57 |
| 0.2 | 0.5 | 200 | 0.4997 | 0.0142 | 2.22 | 0.5001 | 0.0103 | 1.67 |
| 0.2 | 0.5 | 500 | 0.4996 | 0.0087 | 1.40 | 0.4999 | 0.0070 | 1.14 |
| 0.6 | 0.25 | 100 | 0.2372 | 0.0423 | 14.17 | 0.2415 | 0.0408 | 13.31 |
| 0.6 | 0.25 | 200 | 0.2431 | 0.0297 | 9.77 | 0.2454 | 0.0281 | 9.30 |
| 0.6 | 0.25 | 500 | 0.2462 | 0.0174 | 5.77 | 0.2477 | 0.0171 | 5.68 |
| 0.6 | 0.5 | 100 | 0.4862 | 0.0473 | 7.78 | 0.4966 | 0.0407 | 6.38 |
| 0.6 | 0.5 | 200 | 0.4909 | 0.0337 | 5.39 | 0.5003 | 0.0272 | 4.33 |
| 0.6 | 0.5 | 500 | 0.4939 | 0.0190 | 3.23 | 0.5001 | 0.0167 | 2.72 |
| 0.95 | 0.25 | 100 | 0.0545 | 0.0731 | 78.37 | 0.1769 | 0.2478 | 74.87 |
| 0.95 | 0.25 | 200 | 0.0597 | 0.0544 | 76.14 | 0.1919 | 0.1424 | 50.49 |
| 0.95 | 0.25 | 500 | 0.0678 | 0.0343 | 72.86 | 0.2195 | 0.0921 | 31.83 |
| 0.95 | 0.5  | 100 | 0.1340 | 0.0758 | 73.20 | 0.3816 | 0.3372 | 53.59 |
| 0.95 | 0.5  | 200 | 0.1422 | 0.0560 | 71.55 | 0.4521 | 0.1960 | 33.26 |
| 0.95 | 0.5  | 500 | 0.1495 | 0.0362 | 70.10 | 0.4466 | 0.1827 | 21.57 |

Table 4. Predictive Ability and Convergence Rate with Normal Covariate

| $\rho$ | $\delta$ | Series Length | Hybrid Estimation | | | Extended Kalman-Filter | | |
|---|---|---|---|---|---|---|---|---|
| | | | MAPE | rMSE | Conv. Rate | MAPE | rMSE | Conv. Rate |
| 0.2 | 0.25 | 100 | 7.91 | 9.87 | 100 | 7.90 | 9.86 | 100 |
| 0.2 | 0.25 | 200 | 8.07 | 10.06 | 100 | 8.06 | 10.06 | 100 |
| 0.2 | 0.25 | 500 | 8.10 | 10.13 | 100 | 8.10 | 10.13 | 100 |
| 0.2 | 0.5 | 100 | 7.93 | 10.32 | 100 | 7.90 | 10.27 | 100 |
| 0.2 | 0.5 | 200 | 8.08 | 10.54 | 100 | 8.06 | 10.50 | 100 |
| 0.2 | 0.5 | 500 | 8.10 | 10.61 | 100 | 8.09 | 10.60 | 100 |
| 0.6 | 0.25 | 100 | 7.82 | 9.84 | 100 | 7.82 | 9.84 | 100 |
| 0.6 | 0.25 | 200 | 7.97 | 10.02 | 100 | 7.97 | 10.01 | 100 |
| 0.6 | 0.25 | 500 | 8.02 | 10.10 | 100 | 8.01 | 10.10 | 100 |
| 0.6 | 0.5 | 100 | 7.66 | 10.36 | 100 | 7.62 | 10.30 | 100 |
| 0.6 | 0.5 | 200 | 7.79 | 10.52 | 100 | 7.76 | 10.48 | 100 |
| 0.6 | 0.5 | 500 | 7.82 | 10.61 | 100 | 7.80 | 10.60 | 100 |
| 0.95 | 0.25 | 100 | 8.70 | 10.26 | 96 | 7.99 | 9.84 | 99 |
| 0.95 | 0.25 | 200 | 9.28 | 10.67 | 97 | 8.19 | 10.01 | 99 |
| 0.95 | 0.25 | 500 | 9.63 | 10.97 | 98 | 8.26 | 10.09 | 100 |
| 0.95 | 0.5 | 100 | 8.39 | 10.79 | 94 | 7.65 | 10.31 | 99 |
| 0.95 | 0.5 | 200 | 8.84 | 11.24 | 96 | 7.82 | 10.50 | 97 |
| 0.95 | 0.5 | 500 | 9.18 | 11.56 | 98 | 7.87 | 10.56 | 99 |

*Covariate: Uniform(0,1)*

The two estimation procedures are compared with covariates generated from uniform distribution in Tables 5 to 7. Similar with the case of normal covariates, the two estimation procedures are comparable. Convergence rates of both algorithms are high whether the series is stationarity or approaching near non-stationarity. For stationary series, the hybrid method produces estimates and predictive ability comparable to the Extended Kalman-Filter. Both methods though, fail to predict the time series well when the time series is near non-stationarity.

Similarity in results for normal and uniform covariates are not associated with the distribution itself, but rather with the mean level of the covariates. With a standard normal and a standard uniform distribution, the mean levels are close to 0. Given a covariate effect $\delta$ (0.25 or 0.5 in the simulation settings), the effect of the explanatory variable to push away the mean level of the series from its reversion point is minimal.

Table 5. Estimates for $\rho$, Standard Error and Relative Bias with Uniform Covariate

| $\rho$ | $\delta$ | Series Length | Hybrid Estimation | | | Extended Kalman-Filter | | |
|---|---|---|---|---|---|---|---|---|
| | | | Estimate | St. Err. | Relative Bias | Estimate | St. Err. | Relative Bias |
| 0.2 | 0.25 | 100 | 0.1849 | 0.0925 | 38.26 | 0.1837 | 0.0930 | 38.44 |
| 0.2 | 0.25 | 200 | 0.1904 | 0.0621 | 24.14 | 0.1906 | 0.0624 | 24.49 |
| 0.2 | 0.25 | 500 | 0.1952 | 0.0368 | 14.47 | 0.1954 | 0.0362 | 14.12 |
| 0.2 | 0.5 | 100 | 0.1977 | 0.0671 | 27.21 | 0.1978 | 0.0638 | 25.86 |
| 0.2 | 0.5 | 200 | 0.1974 | 0.0432 | 16.71 | 0.1977 | 0.0415 | 16.08 |
| 0.2 | 0.5 | 500 | 0.1975 | 0.0263 | 10.17 | 0.1971 | 0.0256 | 9.99 |
| 0.6 | 0.25 | 100 | 0.5588 | 0.0852 | 12.61 | 0.5652 | 0.0859 | 12.41 |
| 0.6 | 0.25 | 200 | 0.5802 | 0.0556 | 7.73 | 0.5845 | 0.0561 | 7.65 |
| 0.6 | 0.25 | 500 | 0.5906 | 0.0352 | 4.78 | 0.5924 | 0.0349 | 4.71 |
| 0.6 | 0.5 | 100 | 0.5726 | 0.0716 | 10.09 | 0.5764 | 0.0694 | 9.71 |
| 0.6 | 0.5 | 200 | 0.5848 | 0.0494 | 6.81 | 0.5886 | 0.0491 | 6.63 |
| 0.6 | 0.5 | 500 | 0.5929 | 0.0297 | 4.05 | 0.5945 | 0.0295 | 4.00 |
| 0.95 | 0.25 | 100 | 0.8695 | 0.0565 | 8.55 | 0.8953 | 0.0563 | 6.11 |
| 0.95 | 0.25 | 200 | 0.9095 | 0.0334 | 4.39 | 0.9249 | 0.0328 | 3.15 |
| 0.95 | 0.25 | 500 | 0.9338 | 0.0156 | 1.86 | 0.9403 | 0.0155 | 1.41 |
| 0.95 | 0.5 | 100 | 0.8741 | 0.0535 | 8.02 | 0.8987 | 0.0539 | 5.73 |
| 0.95 | 0.5 | 200 | 0.9116 | 0.0319 | 4.17 | 0.9266 | 0.0312 | 2.97 |
| 0.95 | 0.5 | 500 | 0.9335 | 0.0161 | 1.88 | 0.9400 | 0.0160 | 1.46 |

**Table 6. Estimates for Covariate Effect δ, Standard Error, and Relative Bias with Uniform Covariate**

| ρ | δ | Series Length | Hybrid Estimation | | | Extended Kalman-Filter | | |
|---|---|---|---|---|---|---|---|---|
| | | | Estimate | St. Err. | Relative Bias | Estimate | St. Err. | Relative Bias |
| 0.2 | 0.25 | 100 | 0.2451 | 0.0466 | 14.78 | 0.2460 | 0.0463 | 14.47 |
| 0.2 | 0.25 | 200 | 0.2460 | 0.0331 | 10.38 | 0.2467 | 0.0327 | 10.09 |
| 0.2 | 0.25 | 500 | 0.2488 | 0.0215 | 6.92 | 0.2491 | 0.0213 | 6.92 |
| 0.2 | 0.5 | 100 | 0.5005 | 0.0547 | 8.24 | 0.4999 | 0.0522 | 7.95 |
| 0.2 | 0.5 | 200 | 0.4986 | 0.0365 | 5.57 | 0.4984 | 0.0358 | 5.43 |
| 0.2 | 0.5 | 500 | 0.4986 | 0.0237 | 3.77 | 0.4982 | 0.0235 | 3.79 |
| 0.6 | 0.25 | 100 | 0.2279 | 0.0905 | 28.94 | 0.2336 | 0.0942 | 29.69 |
| 0.6 | 0.25 | 200 | 0.2374 | 0.0610 | 19.65 | 0.2414 | 0.0628 | 19.90 |
| 0.6 | 0.25 | 500 | 0.2447 | 0.0417 | 13.62 | 0.2464 | 0.0421 | 13.68 |
| 0.6 | 0.5 | 100 | 0.4757 | 0.1078 | 17.63 | 0.4817 | 0.1088 | 17.40 |
| 0.6 | 0.5 | 200 | 0.4846 | 0.0767 | 12.27 | 0.4907 | 0.0778 | 12.23 |
| 0.6 | 0.5 | 500 | 0.4941 | 0.0492 | 8.00 | 0.4966 | 0.0496 | 7.97 |
| 0.95 | 0.25 | 100 | 0.0500 | 0.2195 | 97.08 | 0.1979 | 0.6664 | 153.37 |
| 0.95 | 0.25 | 200 | 0.0481 | 0.1627 | 87.34 | 0.2078 | 0.5648 | 120.29 |
| 0.95 | 0.25 | 500 | 0.0640 | 0.1194 | 77.57 | 0.2318 | 0.2653 | 83.84 |
| 0.95 | 0.5 | 100 | 0.1276 | 0.1956 | 75.25 | 0.3566 | 0.6658 | 81.18 |
| 0.95 | 0.5 | 200 | 0.1288 | 0.1614 | 74.25 | 0.4049 | 0.5264 | 65.16 |
| 0.95 | 0.5 | 500 | 0.1461 | 0.1168 | 70.79 | 0.4413 | 0.2771 | 44.35 |

**Table 7. Predictive Ability and Convergence Rate with Uniform Covariate**

| ρ | δ | Series Length | Hybrid Estimation | | | Extended Kalman-Filter | | |
|---|---|---|---|---|---|---|---|---|
| | | | MAPE | rMSE | Conv. Rate | MAPE | rMSE | Conv. Rate |
| 0.2 | 0.25 | 100 | 7.52 | 10.49 | 100 | 7.52 | 10.50 | 100 |
| 0.2 | 0.25 | 200 | 7.58 | 10.60 | 100 | 7.58 | 10.60 | 100 |
| 0.2 | 0.25 | 500 | 7.58 | 10.61 | 100 | 7.58 | 10.61 | 100 |
| 0.2 | 0.5 | 100 | 7.04 | 11.20 | 100 | 7.04 | 11.20 | 100 |
| 0.2 | 0.5 | 200 | 7.11 | 11.32 | 100 | 7.11 | 11.32 | 100 |
| 0.2 | 0.5 | 500 | 7.11 | 11.34 | 100 | 7.11 | 11.34 | 100 |
| 0.6 | 0.25 | 100 | 7.51 | 10.48 | 100 | 7.51 | 10.48 | 100 |
| 0.6 | 0.25 | 200 | 7.57 | 10.58 | 100 | 7.57 | 10.58 | 100 |
| 0.6 | 0.25 | 500 | 7.58 | 10.60 | 100 | 7.58 | 10.60 | 100 |
| 0.6 | 0.5 | 100 | 7.03 | 11.20 | 100 | 7.03 | 11.21 | 100 |
| 0.6 | 0.5 | 200 | 7.08 | 11.31 | 100 | 7.08 | 11.31 | 100 |
| 0.6 | 0.5 | 500 | 7.09 | 11.34 | 100 | 7.09 | 11.34 | 100 |
| 0.95 | 0.25 | 100 | 8.26 | 10.75 | 94 | 7.72 | 10.38 | 100 |
| 0.95 | 0.25 | 200 | 8.64 | 11.21 | 97 | 7.78 | 10.59 | 100 |
| 0.95 | 0.25 | 500 | 8.94 | 11.50 | 96 | 7.81 | 10.62 | 100 |
| 0.95 | 0.5 | 100 | 7.64 | 11.51 | 94 | 7.17 | 11.14 | 100 |
| 0.95 | 0.5 | 200 | 7.92 | 11.93 | 96 | 7.22 | 11.32 | 100 |
| 0.95 | 0.5 | 500 | 8.19 | 12.19 | 99 | 7.25 | 11.34 | 100 |

*Covariate: Poisson(5)*

With covariates generated from Poisson distribution, estimates of the parameters of PAR(p) based on the extended Kalman-filter and the hybrid method are presented in Tables 8 and 9 while the indicators of predictive ability are summarized in Table 10. Significant differences in the convergence rate in the two methods are evident in cases with a covariate generated from Poisson distribution. Unlike a standard normal or standard uniform covariate, Poisson covariate produces nonnegative values that are, in general, far away from zero (unless it is zero-inflated as in the case with zero mean). This causes the series to have larger dynamic means than its reversion point. This clearly imposes problems with the convergence of the Extended Kalman-Filter as demonstrated by [1] while the proposed hybrid estimation still exhibits high convergence rates. For stationary series, considering only convergent cases, both estimation methods capture the dynamic behavior of the series as indicated with the standard errors and relative bias. However, due to large dynamic means, even with stationarity, extended Kalman-filtering produces forecasts that are too far from the actual counts resulting to inflated MAPE and rMSE. On the other hand, good predictive ability of the hybrid method is still observed. Both methods are having problems with near non-stationary cases as the PAR(p) model was originally intended for stationary count time series.

These simulation results show the PAR(p) model viewed as an additive model resolves the issue pertaining large dynamic means and performs comparably or better than estimates from the extended Kalman-filter. The primary reason is the relative flexibility of the additive model. Model specification is straightforward enabling the model to be incorporated to more complex dependence structures.

Table 8. Estimates for ρ, Standard Error and Relative Bias with Poisson Covariate

| ρ | δ | Series Length | Hybrid Estimation | | | Extended Kalman-Filter | | |
|---|---|---|---|---|---|---|---|---|
| | | | Estimate | St. Err. | Relative Bias | Estimate | St. Err. | Relative Bias |
| 0.2 | 0.25 | 100 | 0.2008 | 0.0141 | 5.61 | 0.1996 | 0.0127 | 3.88 |
| 0.2 | 0.25 | 200 | 0.1999 | 0.0093 | 3.68 | 0.2005 | 0.0152 | 3.08 |
| 0.2 | 0.25 | 500 | 0.1993 | 0.0061 | 2.44 | 0.1996 | 0.0040 | 1.58 |
| 0.2 | 0.5 | 100 | 0.1999 | 0.0044 | 1.72 | 0.1834 | 0.1254 | 9.09 |
| 0.2 | 0.5 | 200 | 0.1997 | 0.0023 | 0.90 | 0.1941 | 0.0351 | 3.38 |
| 0.2 | 0.5 | 500 | 0.2001 | 0.0015 | 0.60 | 0.2025 | 0.0445 | 4.12 |
| 0.6 | 0.25 | 100 | 0.5979 | 0.0194 | 2.51 | 0.5981 | 0.0289 | 2.24 |
| 0.6 | 0.25 | 200 | 0.5987 | 0.0125 | 1.68 | 0.5966 | 0.0337 | 1.80 |
| 0.6 | 0.25 | 500 | 0.5993 | 0.0086 | 1.11 | 0.5985 | 0.0193 | 1.05 |
| 0.6 | 0.5 | 100 | 0.5998 | 0.0049 | 0.64 | 0.5999 | 0.0024 | 0.31 |
| 0.6 | 0.5 | 200 | 0.5997 | 0.0029 | 0.37 | 0.5931 | 0.0653 | 1.40 |
| 0.6 | 0.5 | 500 | 0.6001 | 0.0018 | 0.23 | 0.5950 | 0.0337 | 1.00 |
| 0.95 | 0.25 | 100 | 0.8975 | 0.0396 | 5.65 | 0.9331 | 0.0318 | 2.68 |
| 0.95 | 0.25 | 200 | 0.9236 | 0.0257 | 2.96 | 0.9435 | 0.0175 | 1.34 |
| 0.95 | 0.25 | 500 | 0.9391 | 0.0139 | 1.42 | 0.9476 | 0.0086 | 0.70 |
| 0.95 | 0.5 | 100 | 0.9347 | 0.0209 | 1.82 | 0.9515 | 0.0092 | 0.46 |
| 0.95 | 0.5 | 200 | 0.9428 | 0.0131 | 0.99 | 0.9517 | 0.0090 | 0.37 |
| 0.95 | 0.5 | 500 | 0.9479 | 0.0043 | 0.40 | 0.9517 | 0.0091 | 0.31 |

Table 9. Estimates for Covariate Effect δ, Standard Error, and Relative Bias with Poisson Covariate

| ρ | δ | Series Length | Hybrid Estimation | | | Extended Kalman-Filter | | |
|---|---|---|---|---|---|---|---|---|
| | | | Estimate | St. Err. | Relative Bias | Estimate | St. Err. | Relative Bias |
| 0.2 | 0.25 | 100 | 0.2501 | 0.0046 | 1.48 | 0.2690 | 0.1915 | 8.76 |
| 0.2 | 0.25 | 200 | 0.2498 | 0.0030 | 0.93 | 0.2588 | 0.1196 | 4.41 |
| 0.2 | 0.25 | 500 | 0.2498 | 0.0019 | 0.61 | 0.2498 | 0.0014 | 0.48 |
| 0.2 | 0.5 | 100 | 0.5001 | 0.0029 | 0.46 | 0.5134 | 0.1222 | 3.24 |
| 0.2 | 0.5 | 200 | 0.4998 | 0.0014 | 0.23 | 0.5449 | 0.2693 | 9.14 |
| 0.2 | 0.5 | 500 | 0.5000 | 0.0009 | 0.14 | 0.5405 | 0.2220 | 8.19 |
| 0.6 | 0.25 | 100 | 0.2488 | 0.0117 | 3.75 | 0.2577 | 0.1046 | 5.74 |

| 0.6 | 0.25 | 200 | 0.2491 | 0.0073 | 2.41 | 0.2678 | 0.1802 | 9.17 |
|---|---|---|---|---|---|---|---|---|
| 0.6 | 0.25 | 500 | 0.2494 | 0.0049 | 1.58 | 0.2582 | 0.1128 | 4.47 |
| 0.6 | 0.5 | 100 | 0.4999 | 0.0061 | 0.99 | 0.5003 | 0.0024 | 0.36 |
| 0.6 | 0.5 | 200 | 0.4992 | 0.0029 | 0.46 | 0.5202 | 0.1948 | 4.36 |
| 0.6 | 0.5 | 500 | 0.4997 | 0.0019 | 0.30 | 0.5250 | 0.1641 | 5.16 |
| 0.95 | 0.25 | 100 | 0.1172 | 0.0329 | 53.10 | 0.2299 | 0.1016 | 32.07 |
| 0.95 | 0.25 | 200 | 0.1341 | 0.0260 | 46.36 | 0.2380 | 0.0565 | 17.65 |
| 0.95 | 0.25 | 500 | 0.1472 | 0.0164 | 41.11 | 0.2447 | 0.0313 | 10.02 |
| 0.95 | 0.5 | 100 | 0.3765 | 0.0722 | 24.72 | 0.5060 | 0.0301 | 4.38 |
| 0.95 | 0.5 | 200 | 0.4079 | 0.0486 | 18.43 | 0.5035 | 0.0252 | 3.26 |
| 0.95 | 0.5 | 500 | 0.4320 | 0.0217 | 13.59 | 0.5024 | 0.0164 | 1.83 |

**Table 10. Predictive Ability and Convergence Rate with Poisson Covariate**

| $\rho$ | $\delta$ | Series Length | Hybrid Estimation | | | Extended Kalman-Filter | | |
|---|---|---|---|---|---|---|---|---|
| | | | MAPE | rMSE | Conv. Rate | MAPE | rMSE | Conv. Rate |
| 0.2 | 0.25 | 100 | 4.28 | 20.11 | 100 | 6.075E9 | 9.056E11 | 92 |
| 0.2 | 0.25 | 200 | 4.28 | 20.11 | 100 | 3.221E5 | 2.531E7 | 86 |
| 0.2 | 0.25 | 500 | 4.28 | 20.26 | 100 | 4.27 | 20.22 | 85 |
| 0.2 | 0.5 | 100 | 2.28 | 49.47 | 100 | 7857.18 | 1.606E7 | 37 |
| 0.2 | 0.5 | 200 | 2.29 | 48.17 | 100 | 2.271E6 | 1.757E9 | 36 |
| 0.2 | 0.5 | 500 | 2.26 | 50.42 | 100 | 8.534E5 | 5.065E9 | 30 |
| 0.6 | 0.25 | 100 | 4.07 | 20.04 | 100 | 1.553E5 | 6.383E6 | 90 |
| 0.6 | 0.25 | 200 | 4.09 | 20.18 | 100 | 2.102E7 | 1.803E9 | 96 |
| 0.6 | 0.25 | 500 | 4.08 | 20.25 | 100 | 1.081E6 | 7.513E6 | 90 |
| 0.6 | 0.5 | 100 | 1.96 | 51.52 | 100 | 1.93 | 47.88 | 46 |
| 0.6 | 0.5 | 200 | 1.93 | 51.07 | 98 | 3.533E9 | 1.967E13 | 46 |
| 0.6 | 0.5 | 500 | 1.92 | 49.95 | 100 | 82323.16 | 3.782E8 | 44 |
| 0.95 | 0.25 | 100 | 4.12 | 20.67 | 93 | 3.95 | 20.00 | 100 |
| 0.95 | 0.25 | 200 | 4.18 | 21.07 | 96 | 3.95 | 20.21 | 98 |
| 0.95 | 0.25 | 500 | 4.25 | 21.41 | 98 | 3.96 | 20.27 | 100 |
| 0.95 | 0.5 | 100 | 1.97 | 64.84 | 89 | 1.70 | 51.38 | 63 |
| 0.95 | 0.5 | 200 | 1.91 | 66.20 | 94 | 1.73 | 52.35 | 57 |
| 0.95 | 0.5 | 500 | 1.83 | 61.92 | 96 | 1.73 | 51.61 | 41 |

## 5. ROBUSTNESS TO TEMPORARY STRUCTURAL CHANGE

Time-dependent count processes are oftentimes vulnerable to short-term random shocks and temporary structural changes triggered by external factors influencing the phenomenon they ought to characterize. These changes causing disturbances into the behavior of the series imposes grave issues in modeling. Suppose that some external factors caused a temporary structural change, affecting the autocorrelation pattern of the count series. With temporary rather than permanent change, we expect the time series to revert to its original behavior, this necessitates estimator that are unaffected by the change. This section examines the implication of temporary structural change to the characteristics of the estimates produced by the hybrid estimation procedure.

We simulate scenarios that induces temporary structural change on the PAR model. The temporary structural change is introduced in the middle portion of the series which persists only until some portion in the series. The simulation study focuses on the capability of the hybrid estimation procedure to estimate the model given a sudden shift in the autocorrelation pattern. A shift in mean is not included since it will mimic the behavior of trend non-stationarity, the effect of which has been analyzed in the previous section. We summarized all the scenarios simulated to study the effect of temporary structural change on the hybrid estimation procedure in Table 11.

**Table 11. Summary of Scenarios for Temporary Structural Change**

| Temporary Change in Autocorrelation | Position/Length of Structural Change | Series Length |
|---|---|---|
| From $\rho = 0.20$ to $\rho = 0.60$<br>From $\rho = 0.20$ to $\rho = 0.95$<br>From $\rho = 0.60$ to $\rho = 0.95$<br>From $\rho = 0.80$ to $\rho = 0.95$ | middle 10%<br>middle 25% | 100<br>300<br>500 |

The presence of temporary structural change does not affect the convergence rate and the number of iterations required until convergence of the hybrid algorithm. In the simulated scenarios, the average number of iterations is at most 6 with convergence rate of at least 99.50%. Summary of estimates, standard errors, percent biases and predictive ability in terms of MAPE and MSE are presented Table 12.

The estimate of the autoregressive parameters are affected by the magnitude of the temporary structural change, extent of persistence of the temporary change, and length of the time series. Provided that the temporary structural change does not drive the time series towards non-stationarity, the standard error and relative bias of estimates from the hybrid method are within manageable level and can be controlled further with minimal extent and persistence of the temporary structural change, and for longer time series, see Table 12 for details.

Table 12. Estimates of the Autoregressive Parameter $\rho$, Standard Error and Relative Bias in the Presence of Temporary Structural Change

| Temporary Structural Change Induced | Position and Length of Structural Change | Series Length | Estimate | St. Err. | Relative Bias |
|---|---|---|---|---|---|
| $\rho = 0.20 \rightarrow \rho = 0.60$ | middle 10% | 100 | 0.2132 | 0.0935 | 36.52 |
| $\rho = 0.20 \rightarrow \rho = 0.60$ | middle 10% | 300 | 0.2231 | 0.0563 | 24.12 |
| $\rho = 0.20 \rightarrow \rho = 0.60$ | middle 10% | 500 | 0.2230 | 0.0474 | 21.10 |
| $\rho = 0.20 \rightarrow \rho = 0.60$ | middle 25% | 100 | 0.2575 | 0.0965 | 44.97 |
| $\rho = 0.20 \rightarrow \rho = 0.60$ | middle 25% | 300 | 0.2698 | 0.0563 | 37.21 |
| $\rho = 0.20 \rightarrow \rho = 0.60$ | middle 25% | 500 | 0.2718 | 0.0465 | 37.32 |
| $\rho = 0.20 \rightarrow \rho = 0.95$ | middle 10% | 100 | 0.2999 | 0.1259 | 65.47 |
| $\rho = 0.20 \rightarrow \rho = 0.95$ | middle 10% | 300 | 0.3459 | 0.0897 | 74.43 |
| $\rho = 0.20 \rightarrow \rho = 0.95$ | middle 10% | 500 | 0.3652 | 0.0794 | 82.87 |
| $\rho = 0.20 \rightarrow \rho = 0.95$ | middle 25% | 100 | 0.4376 | 0.1423 | 121.21 |
| $\rho = 0.20 \rightarrow \rho = 0.95$ | middle 25% | 300 | 0.5126 | 0.0890 | 156.28 |
| $\rho = 0.20 \rightarrow \rho = 0.95$ | middle 25% | 500 | 0.5256 | 0.0788 | 162.82 |
| $\rho = 0.60 \rightarrow \rho = 0.95$ | middle 10% | 100 | 0.6118 | 0.0796 | 10.82 |
| $\rho = 0.60 \to \rho = 0.95$ | middle 10% | 300 | 0.6549 | 0.0513 | 10.12 |
| $\rho = 0.60 \to \rho = 0.95$ | middle 10% | 500 | 0.6657 | 0.0434 | 11.54 |
| $\rho = 0.60 \to \rho = 0.95$ | middle 25% | 100 | 0.6794 | 0.0831 | 16.14 |
| $\rho = 0.60 \to \rho = 0.95$ | middle 25% | 300 | 0.7363 | 0.0521 | 22.72 |
| $\rho = 0.60 \to \rho = 0.95$ | middle 25% | 500 | 0.7479 | 0.0446 | 24.66 |
| $\rho = 0.80 \to \rho = 0.95$ | middle 10% | 100 | 0.7649 | 0.0694 | 7.50 |
| $\rho = 0.80 \to \rho = 0.95$ | middle 10% | 300 | 0.8087 | 0.0386 | 4.01 |
| $\rho = 0.80 \to \rho = 0.95$ | middle 10% | 500 | 0.8172 | 0.0295 | 3.53 |
| $\rho = 0.80 \to \rho = 0.95$ | middle 25% | 100 | 0.7905 | 0.0710 | 7.21 |
| $\rho = 0.80 \to \rho = 0.95$ | middle 25% | 300 | 0.8414 | 0.0381 | 5.90 |
| $\rho = 0.80 \to \rho = 0.95$ | middle 25% | 500 | 0.8497 | 0.0297 | 6.39 |

Table 13. Predictive Ability in the Presence of Temporary Structural Change

| Temporary Structural Change Induced | Position and Length of Structural Change | Series Length | MAPE | rMSE |
|---|---|---|---|---|
| $\rho = 0.20 \to \rho = 0.60$ | middle 10% | 100 | 8.52 | 10.71 |
| $\rho = 0.20 \to \rho = 0.60$ | middle 10% | 300 | 8.56 | 10.86 |
| $\rho = 0.20 \to \rho = 0.60$ | middle 10% | 500 | 8.58 | 10.85 |
| $\rho = 0.20 \to \rho = 0.60$ | middle 25% | 100 | 8.98 | 11.30 |
| $\rho = 0.20 \to \rho = 0.60$ | middle 25% | 300 | 9.09 | 11.55 |
| $\rho = 0.20 \to \rho = 0.60$ | middle 25% | 500 | 9.12 | 11.57 |
| $\rho = 0.20 \to \rho = 0.95$ | middle 10% | 100 | 9.75 | 12.53 |
| $\rho = 0.20 \to \rho = 0.95$ | middle 10% | 300 | 10.10 | 13.14 |
| $\rho = 0.20 \to \rho = 0.95$ | middle 10% | 500 | 10.33 | 13.37 |
| $\rho = 0.20 \to \rho = 0.95$ | middle 25% | 100 | 11.83 | 14.80 |
| $\rho = 0.20 \to \rho = 0.95$ | middle 25% | 300 | 12.56 | 15.78 |
| $\rho = 0.20 \to \rho = 0.95$ | middle 25% | 500 | 12.71 | 15.92 |
| $\rho = 0.60 \to \rho = 0.95$ | middle 10% | 100 | 8.32 | 10.45 |
| $\rho = 0.60 \to \rho = 0.95$ | middle 10% | 300 | 8.55 | 10.77 |
| $\rho = 0.60 \to \rho = 0.95$ | middle 10% | 500 | 8.58 | 10.80 |
| $\rho = 0.60 \to \rho = 0.95$ | middle 25% | 100 | 8.86 | 10.99 |
| $\rho = 0.60 \to \rho = 0.95$ | middle 25% | 300 | 9.16 | 11.39 |
| $\rho = 0.60 \to \rho = 0.95$ | middle 25% | 500 | 9.17 | 11.39 |
| $\rho = 0.80 \to \rho = 0.95$ | middle 10% | 100 | 7.97 | 9.95 |
| $\rho = 0.80 \to \rho = 0.95$ | middle 10% | 300 | 8.17 | 10.21 |
| $\rho = 0.80 \to \rho = 0.95$ | middle 10% | 500 | 8.20 | 10.24 |
| $\rho = 0.80 \to \rho = 0.95$ | middle 25% | 100 | 8.14 | 10.04 |
| $\rho = 0.80 \to \rho = 0.95$ | middle 25% | 300 | 8.37 | 10.33 |
| $\rho = 0.80 \to \rho = 0.95$ | middle 25% | 500 | 8.38 | 10.36 |

As indicated in Table 13, predictive ability of the hybrid estimation method is robust to the presence of temporary structural change. Predictive ability is also robust to the length of the time series as well as the extent and persistence of the temporary structural change. Observe further that predictive ability is still good even in temporary structural change that drives the time series towards near non-stationarity.

## 6. GENERALIZATION OF POISSON AUTOREGRESSIVE MODEL TO MULTIPLE TIME SERIES

We also take advantage of the additivity of the PAR(p) model and postulate its extension to a multiple time series in [13]. Consider N uncorrelated time series $\{Y_{1,t}, Y_{2,t}, \ldots, Y_{N,t}\}, t = 1,2,\ldots,T$ that exhibit similar dynamic patterns. [13] formulated a multiple time series model where each series share a common autoregressive parameter $\phi$ with random effect component as manifestation of the unique behavior of each time series, i.e.,

$$Y_{i,t} = \phi Y_{i,t-1} + \lambda_i + \varepsilon_{i,t} \tag{13}$$

where $\quad \lambda_i \sim (\mu_i, \sigma^2_{\lambda_i})$ and $\varepsilon_{i,t} \sim (0, \sigma^2_\varepsilon)$ \quad (14)

Suppose that N uncorrelated count time series $\{Y_{1,t}, Y_{2,t}, \ldots, Y_{N,t}\}, t = 1,2,\ldots,T$ are given. The postulated model is summarized as follows:

$$Y_{i,t} \sim \text{Poisson}(m_t) \text{ i.e., } P(Y_{i,t} = y | m_{i,t}) = \frac{m_{i,t}^y e^{-m_{i,t}}}{y!} \tag{15}$$

$$m_{i,t} = \rho Y_{i,t-1} + (1-\rho)\exp(\delta_{i,0} + X_{i,t}`\boldsymbol{\delta}) \tag{16}$$

where $Y_{i,t}$ – count response of $i^{\text{th}}$ series at time t

$m_{i,t}$ – dynamic mean of the Poisson data generating process of $i^{th}$ series at time $t$

$\rho$ – common autoregressive parameters

$\boldsymbol{X}_{i,t}$ – vector of covariates of $i^{th}$ series at time $t$

$\boldsymbol{\delta}$ – common vector of coefficients of the covariates

$\delta_{i,o}$ – mean-reversion point of the count series of $i^{th}$ series

Equation (16) is an analogous transition equation to the multiple time series model in Equation (13) where the individual behavior component $\lambda_i$ is expressed as $(1-\rho)\exp(\delta_{i,0} + \boldsymbol{X}_{i,t}`\boldsymbol{\delta})$. There are two differences in these individual effect specifications: 1) the individual effect of the former is a random component while the latter is fixed. 2) a random constant summarizes entirely the differences between each series for the former formulation while the latter specifies the differences via varying mean-reversion point while allowing for the contribution of the covariates. Most stationary count process usually exhibit a fixed stabilization level due to some external factors governing the generation of counts. In epidemics, prevalence of certain diseases in an area is a function of the endowments or risk factors present in the community, e.g., population. Prevalence of the disease cannot exceed the area's population. Therefore, instead of being random, we incorporated a varying but fixed mean-reversion point for each series. The benefit of such formulation that that it enables the multiple PAR model simplify to the original PAR model when N = 1.

*Estimation Procedure of the Multiple PAR(p) Model*

It is clear from Equation (16) that the postulated multiple count time series exhibits additivity. To take advantage of this, rather than the tedious deriving its likelihood function, we estimate the postulated model through the backfitting algorithm based on the proposed hybrid estimation procedure for single PAR(p) model embedded in the estimation procedure of [13].

Equation (16) is estimated as follows:

*Step1*: For each of the N count series, ignore the common autoregressive term and fit $Y_{i,t}$ with covariates using Poisson Regression to obtain $\hat{\delta}_{i,o}^{(0)}$ and $\widehat{\boldsymbol{\delta}}_i^{(0)}$. The varying mean-reversion points are estimated by $\hat{\delta}_{i,o}^{(0)}$ while the common covariate effects are estimated by $\hat{\boldsymbol{\delta}}^{(0)} = \frac{\sum_{i=1}^{N} \widehat{\boldsymbol{\delta}}_i^{(0)}}{N}$, i.e., the average of all estimated covariate effects for each series.

*Step2*: Compute residuals from $R_{i,t}^{(1)} = Y_{i,t} - exp\,(\hat{\delta}_{i,o}^{(0)} + \boldsymbol{X}_{i,t}`\hat{\boldsymbol{\delta}}^{(0)})$.

*Step3*: For each of the N count series, fit $R_{i,t}^{(1)}$ as a cubic smoothing spline function of $R_{i,t-i}^{(1)}$ to generate an estimator of the first component denoted by $\widehat{f}_1(Y_{i,t-1})$. The estimator for the autoregressive parameter for each series is given by $\hat{\rho}_i^{(0)} = \left(\sum_{t=2}^{T} \frac{\partial \widehat{f}_1(Y_{i,t-1})}{\partial R_{i,t-i}^{(1)}}\right) \Big/ (T-1)$.

*Step4*: Perform bootstrap resampling on $\hat{\rho}_i^{(0)}$ to obtain $\hat{\rho}^{BS^{(0)}}$, the bootstrap estimate of the shared autoregressive parameter $\rho$.

*Step5*: Compute new residuals from $R_{i,t}^{(2)} = Y_{i,t} - \hat{\rho}^{BS^{(0)}} Y_{i,t-1}$.

For j = 1, 2, 3, … where *j* is the index of iteration, initiate the residuals with $R_{i,t}^{(j)} = R_{i,t}^{(2)}$.

*Step6*: For each of the N count series, fit $R_{i,t}^{(j)}$ with covariates using Poisson Regression to obtain $\hat{\delta}_{i,o}^{(j)}$ and $\widehat{\boldsymbol{\delta}}_i^{(j)}$. The varying mean-reversion points are estimated by $\hat{\delta}_{i,o}^{(j)}$ while the common covariate effects are estimated by $\hat{\boldsymbol{\delta}}^{(j)} = \frac{\sum_{i=1}^{N} \widehat{\boldsymbol{\delta}}_i^{(j)}}{N}$, i.e., the average of all estimated covariate effects for each series.

*Step7*: Compute new residual as $R_{i,t}^{(j)} = Y_{i,t} - exp(\hat{\delta}_{i,o}^{(j)} + X_{i,t}`\hat{\pmb{\delta}}^{(j)})$.

*Step8*: For each of the N count series, fit $R_{i,t}^{(j)}$ as a cubic smoothing spline function of $R_{i,t-i}^{(j)}$ to have an estimator of the first component denoted by $\hat{f}_1(Y_{i,t-1})$. The estimator for the autoregressive parameter for each series is given by $\hat{\rho}_i^{(j)} = \frac{\left(\sum_{t=2}^{T} \frac{\partial \hat{f}_1(Y_{i,t-1})}{\partial R_{i,t-i}^{(j)}}\right)}{(T-1)}$.

*Step9*: Perform bootstrap resampling on $\hat{\rho}_i^{(j)}$ to obtain $\hat{\rho}^{BS^{(j)}}$, the bootstrap estimate of the shared autoregressive parameter $\rho$.

*Step10*: Define new residual as $R_{i,t}^{(j+1)} = Y_{i,t} - \hat{\rho}^{BS^{(j)}} Y_{i,t-1}$.

Iterate from *Step5*, defining residuals using the updated estimates for $\rho$ and $\pmb{\delta}$ until the convergence criterion is achieved, e.g., changes in all parameter estimates are less than 0.0001%.

## *Simulation Results*

We simulated similar scenarios for the multiple count time series data generating process and implemented the estimation procedure. Our interest in the simulation are the relationship between the number of count series, length of each series and variability of the mean-reversion points, and the ability of the hybrid estimation procedure to characterize the postulated model. The advantages of using the hybrid estimation observed in Section 4 are still evident, specifically, in effect of stationarity and distributional assumption on the covariates. We only show results of estimating multiple PAR(p=1) model with common autoregressive parameter $\rho = 0.6$ and common covariate effect $\delta = 0.5$ assuming a standard normal covariate.

**Table 14. Parameter Estimates, Standard Errors, Relative Bias, MAPE and rMSE in Multiple PAR(p)**

| N | T | St. Dev. Varying Means | $\hat{\rho}$ | $se(\hat{\rho})$ | Relative Bias($\hat{\rho}$) | $\hat{\delta}$ | $se(\hat{\delta})$ | Relative Bias($\hat{\delta}$) | MAPE | rMSE |
|---|---|---|---|---|---|---|---|---|---|---|
| 10 | 50 | 5 | 0.5434 | 0.0302 | 9.62 | 0.4387 | 0.0321 | 12.58 | 8.09 | 10.75 |
| 20 | 50 | 5 | 0.5416 | 0.0223 | 9.74 | 0.4355 | 0.0227 | 12.91 | 8.09 | 10.76 |
| 50 | 50 | 5 | 0.5434 | 0.0137 | 9.44 | 0.4369 | 0.0143 | 12.61 | 8.08 | 10.73 |
| 10 | 100 | 5 | 0.5696 | 0.0231 | 5.48 | 0.4632 | 0.0249 | 7.67 | 7.94 | 10.68 |
| 20 | 100 | 5 | 0.5691 | 0.0155 | 5.17 | 0.4624 | 0.0172 | 7.55 | 7.93 | 10.68 |
| 50 | 100 | 5 | 0.5690 | 0.0097 | 5.17 | 0.4623 | 0.0104 | 7.55 | 7.93 | 10.67 |
| 10 | 50 | 10 | 0.5446 | 0.0307 | 9.42 | 0.4400 | 0.0312 | 12.23 | 8.10 | 10.76 |
| 20 | 50 | 10 | 0.5434 | 0.0220 | 9.47 | 0.4375 | 0.0226 | 12.54 | 8.10 | 10.75 |
| 50 | 50 | 10 | 0.5424 | 0.0143 | 9.60 | 0.4362 | 0.0149 | 12.77 | 8.10 | 10.74 |
| 10 | 100 | 10 | 0.5705 | 0.0222 | 5.22 | 0.4639 | 0.0242 | 7.45 | 7.95 | 10.68 |
| 20 | 100 | 10 | 0.5685 | 0.0149 | 5.27 | 0.4620 | 0.0162 | 7.62 | 7.95 | 10.67 |
| 50 | 100 | 10 | 0.5693 | 0.0096 | 5.11 | 0.4626 | 0.0103 | 7.49 | 7.95 | 10.66 |
| 10 | 50 | 20 | 0.5416 | 0.0312 | 9.92 | 0.4373 | 0.0325 | 12.82 | 8.20 | 10.80 |
| 20 | 50 | 20 | 0.5418 | 0.0234 | 9.72 | 0.4361 | 0.0236 | 12.81 | 8.21 | 10.79 |
| 50 | 50 | 20 | 0.5438 | 0.0147 | 9.36 | 0.4374 | 0.0154 | 12.53 | 8.19 | 10.74 |
| 10 | 100 | 20 | 0.5700 | 0.0225 | 5.27 | 0.4630 | 0.0244 | 7.62 | 8.02 | 10.71 |
| 20 | 100 | 20 | 0.5682 | 0.0162 | 5.33 | 0.4610 | 0.0175 | 7.81 | 8.03 | 10.70 |
| 50 | 100 | 20 | 0.5690 | 0.0098 | 5.17 | 0.4618 | 0.0105 | 7.64 | 8.06 | 10.67 |

The relative bias in Table 14 indicates the ability of the hybrid estimation procedure to estimate the parameters of the multiple stationary count time series model. This is further robust to some parameters of the data generating process. Standard errors of estimators and relative biases are relatively low even for shorter time series and improves as length increases. Predictive ability is further good provided that the count time series are stationary.

## 7.     MONTHLY COUNT DAILY INCREASE IN STOCK INDICES

Asian stock markets gained interest both among investors and financial analysts because of the growing number or emerging economies in the region. It has also been identified to be the origin of some financial crisis causing contagion in the global market. What contributes to the growth of Asian stock markets? With data from 12 emerging Asia-Pacific economies for the period 1990-2015, [4] noted that fiscal consolidation attempt contributed on stock prices movement. [12] also noted from 9 Asian developing economies with data from 2001-2017 the mutual causality between fund flows and the macroeconomic indicators. While global factors triggered the initiation of a financial crisis, [7] observed that financial factors are more important drivers of stock returns.

To illustrate the hybrid estimation of PAR(p) model in multiple count time series data, we consider four Asian stock markets: Heng Seng, Nikkei, PSE(Philippines), and Shanghai indices.  We opt not to analyze the index per se since it has a more complicated dependence structure common in high frequency data like severe volatility and other nonlinear dynamics. We counted the number of days that the index increased (closing is higher than the opening) in a month as the time series data. A continually growing market for the whole month will register a count of 23 (approximate number of trading days in a month) and no growth for the whole month implies the count should be zero.  Daily data from 2011-2016 was aggregated and used in estimating the model. Shanghai is fast growing with minimum of 12 days increasing index in a month for the period and average of 20.09, implying that it grows for 20 trading days in a month. Heng Seng has been growing erratically with average of 9.78 and a coefficient of variation of 23.25% (Shanghai had a coefficient of variation of 12.01%). The year 2015 also registered the most erratic growth among the markets with coefficient of variation of 40.17% on the count indicator.

Average Yen-Dollar exchange rate for the month was considered as the covariate. Overall MAPE of the model is 24.97% and rMSE is 3.21. The mean absolute deviation (MAD) was also computed at 2.71, implying that the difference between the actual and predicted count differs by not more than 3 in a month. Similar MAD were computed for the for markets (Heng Seng-2.62; NIKKEI-2.91; PSE-2.73; Shanghai-2.58). The common autoregressive coefficient is 0.0471, indicating a less similar movement of growth among the four markets. There are individual peculiar growth-stimulus among the markets as indicated in their respective intercepts (Heng Seng-2.62; NIKKEI-2.15; PSE-2.74; Shanghai-3.04). The intercepts indicates that Shanghai has higher likelihood to grow while NIKKEI has the least likelihood to grow among the four markets. Yen-Dollar exchange rate as a covariate yield a coefficient of -0.0018 implying that stronger Yen to a Dollar can lead to growth of the four markets.

## 8. CONCLUSIONS

Stylized facts about count time series are often ignored by modelers, resulting to invalid inferences. To account many features of count time series (discreteness, stationarity, mean-reverting and possible overdispersion), [2] developed the Poisson autoregressive PAR(p) model and was estimated through the extended Kalman-filter. This estimation method however suffers from parameter- and state-estimation problems whenever the dynamic means are large. To resolve this issue, PAR(p) model is viewed as an additive model and is estimated through a hybrid estimation in the backfitting framework. Simulation studies shows comparable estimates and predictive abilities of models estimated through the extended Kalman-filter and the hybrid method for stationary data with lower means. The hybrid method however has been shown to work well with large dynamic means when the extended Kalman-filter suffers.

The hybrid estimation procedure for the PAR model is also fairly robust in the presence of temporary structural change. Furthermore, additive nature of PAR(p) facilitates the extension of the model into multiple time series. The hybrid estimation method also mitigates the cumbersome nature of the likelihood function for the multiple time series PAR(p) model needed in extended Kalman-filter.